\documentclass[fleqn,10pt]{olplainarticle}

\title{Driven Dissipative Soliton Resonance}

\author[1]{Vladimir L. Kalashnikov}
\author[1]{Alexander Rudenkov}
\author[2,3]{Evgeni Sorokin}
\author[1,3]{Irina T. Sorokina}
\affil[1]{Department of Physics, Norwegian University of Science and Technology, 7491 Trondheim, Norway}
\affil[2]{Institut f{\"u}r Photonik, TU Wien,  Gu{\ss}hausstra{\ss}e 27/387, A-1040 Vienna, Austria}
\affil[3]{ATLA lasers AS, Richard Birkelands vei 2B, 7034 Trondheim, Norway}


\begin{abstract}
We investigate the enhancement of the dissipative soliton energy scalability by the injection of a low-power single-mode seed synchronized with a chirped-pulse oscillator round-trip. It is demonstrated that a threshold-like transition to multiple-pulse generation limits the maximum energies of dissipative solitons, in agreement with the thermodynamic interpretation of a strongly chirped pulse stability. We show that there are ``islands'' of instability within a stability range of energies which result from stochastic resonance between ``internal modes'' of soliton and quantum noise of the ``basin''. The transition to multiple-pulsing can be suppressed in a system driven by a comparatively low-power seed. However, seed power growth increases the mode-locking energy threshold and produces ``islands'' of instability as the dissipative soliton energy rises. 
\end{abstract}

\begin{document}

\flushbottom
\maketitle
\thispagestyle{empty}

\section*{Introduction}\label{sec:intro}

Recent advances in ultrafast laser technology have enabled mode-locked oscillators to reach unprecedented regimes, from tabletop femtosecond lasers with peak powers exceeding petawatts to compact mid-infrared systems delivering microjoule pulses. At the heart of these achievements lies the concept of dissipative solitons – stable ultrashort pulses that persist due to a delicate balance between dispersion and nonlinearity, as well as between gain and loss \cite{grelu2012dissipative}. Such solitons exemplify self-organization in driven dissipative systems, appearing not only in mode-locked lasers but also in microresonator-based frequency combs as dissipative Kerr solitons \cite{doi:10.1126/science.aan8083}.

In all these platforms, a continuous external drive (e.g., a pump laser) or self-regulating gain provides the energy influx, enabling solitary waves that behave analogously to localized condensates of light. Notably, the formation of a narrow “spectral spike” in a soliton’s spectrum under energy scaling has been likened to a Bose–Einstein condensate (BEC) in frequency space. However, unlike true BECs, where condensation reduces entropy, this spectral concentration in dissipative solitons increases the number of accessible microstates and entropy, heralding a transition toward turbulence and multipulse breakup \cite{kalashnikov2025energy}. This key distinction underpins a thermodynamic perspective on dissipative soliton resonance (DSR) \cite{grelu2010dissipative}: as one drives a mode-locked laser toward higher pulse energies, the soliton broadens asymptotically. 

However, beyond a critical point, the single-soliton state becomes thermodynamically disfavored. The system enters a nonequilibrium “negative temperature” regime in which generating multiple soliton pulses (an energy-unscalable state) maximizes entropy, thus limiting further energy scaling. In practical terms, this manifests as multipulse instabilities: beyond a threshold, increasing pump power tends to spawn additional pulses or soliton molecules within the cavity, rather than a single pulse of higher energy. Experiments and modeling show pronounced hysteresis and multistability in these regimes \cite{komarov2013competition,chowdhury2018multipulse}. A laser may support either a single high-energy soliton or multiple lower-energy solitons under the same conditions, depending on the system’s history and perturbations. These unresolved challenges, including the stability of dissipative solitons (DSs) and energy scalability, motivate a more in-depth theoretical and experimental investigation.

One promising approach to addressing these questions is through the external seeding of mode-locked oscillators. By injecting a low-power continuous-wave (CW) or single-frequency signal into a mode-locked laser, one can perturb the intracavity field in a controlled way. Recent studies have demonstrated that such seeding enables one-by-one manipulation of cavity solitons: for example, injecting a weak CW at a suitable wavelength can trigger the birth or annihilation of individual pulses, allowing fine tuning of the pulse count and repetition rate \cite{korobko2023birth}.

In essence, DS ``feels'' the injected signal through its own ``internal modes'', which can drastically alter the pulse equilibrium. This would allow for the potential to stabilize or destabilize specific soliton configurations through weak driving in a finely controlled manner. They also connect to the broader family of noise–nonlinearity effects: stochastic resonance (SR), where a weak periodic drive is amplified by ambient noise in multistable or threshold systems \cite{RevModPhys.70.223}, and its converse phenomena, stochastic anti-resonance/noise-enhanced stability (SAR), in which a regular drive together with noise suppresses the system’s response or stabilizes a metastable state \cite{kalashnikov2017chaotic}.

In this context, the present work explores driven DSR as a new paradigm for controlling and understanding DS stability. We investigate how a low-power single-mode injection modifies the energy and stability landscape of DSs by resonantly exciting internal soliton modes and leveraging fluctuations induced by the surrounding noise basin. The results open the way for improved control of high-energy mode-locked lasers and offer fresh insights into the universal physics of DSs in lasers, microresonators, and analogous nonlinear systems, including turbulence and BEC.

\section*{Model}\label{sec:model}

As a grounding model, we utilize the following version of the complex cubic-quintic nonlinear Ginzburg-Landau equation (NGLE) \cite{malomed1990kinks,cross1993pattern,kalashnikov2025energy}:
\begin{gather}
 \frac{\partial}{\partial z} a\! \left(z,t\right)=-\sigma a\! \left(z,t\right)+\left(\alpha+i \beta\right) \frac{\partial^{2}}{\partial t^{2}} a\! \left(z,t\right)-
 -i \gamma P\! \left(z,t\right) a\! \left(z,t\right)+\nonumber \\ 
 + \kappa\left(1-\zeta P\! \left(z,t\right)\right) P\! \left(z,t\right) a\! \left(z,t\right)+S(t)+\Gamma(z,t),    \label{eq:NGLE}
\end{gather}

\noindent where $ a\! \left(z,t\right)$ is a laser field amplitude slowly varying with a local time $t$, $z$ is a laser oscillator round-trip number, $\sigma$ is a saturated net-loss coefficient, $\alpha=\left( 2\pi c \Delta \lambda/\lambda^2 \right)^{-2}$, and $\beta$ are the squared inverse spectral filter bandwidth and group-delay dispersion (GDD) coefficients, respectively. $\gamma$ and $\kappa$ are the self-phase (SPM) and self-amplitude (SAM) modulation coefficients, respectively. $\zeta$ is a coefficient characterizing the saturation of self-amplitude modulation. $P \left(z,t\right) =  |a\! \left(z,t\right)|^2$ is a slowly varying power.

The important generalization of an ordinary stochastic NGLE (\ref{eq:NGLE}) is the addition of a coherent localized seed, or forcing term, $S(t)$ \cite{battogtokh1996controlling,chate1999forcing,goyal2012lorentzian}. In our case, it corresponds to a driving single-mode signal at $\lambda$, which is added synchronously with an oscillator period:  

\begin{equation}
    S(t)=\sqrt{A}\cos{(\pi t/2 T_{cav})},
    \label{eq:seed}
\end{equation}

\noindent where $A$ is a seed power, $T_{cav}$ is a cavity period, and temporal window in (\ref{eq:NGLE}) is centered at $t=0$.

To investigate the contribution of quantum noise, we include in Eq. (\ref{eq:NGLE}) an additive complex Gaussian white noise term $\Gamma(z,t)$ \cite{206583,kalashnikov2025energy}:
\begin{gather} \label{eq:noise}
\left\langle \Gamma_m (z)\Gamma_n^* (z') \right\rangle = W \delta_{mn}\delta(z-z'), \nonumber \\ 
\left\langle \Gamma_m (z)\Gamma_n (z') \right\rangle = 0,\\
W = 2 h \nu |\sigma|/T_{cav}.\nonumber 
\end{gather} 

\noindent Here $W$ is a noise power and $\nu$ is a carrier frequency. We assume the DS energy scaling by $T_{cav}$ so that a CW power $P_{av}$ is constant. The energy scaling parameter is a ``CW energy'' $E_{cw} = P_{av} T_{cav}$.

To provide stability, one must take into account the gain saturation, which can be described in the vicinity of the lasing threshold as

\begin{equation}
    \sigma \approx \vartheta (E/E_{cw}-1), \label{sig}
\end{equation}

\noindent where $E=\int_{-\infty }^{\infty }P(z,t)dt$ is a total energy and $\vartheta$ is a ``stiffness'' parameter \cite{kalashnikov2025energy}. 

To be more realistic, we focus on the parameters of a Cr$^{2+}$:ZnS chirped-pulse oscillator (CPO) presented in Table \ref{tab:t1} \cite{Rudenkov}. Eq. (\ref{eq:NGLE}) was simulated numerically by a symmetrical split-step Fourier method on a time window covered by 10$^{16}$ points with 10 fs time interval.

\begin{table}[ht]
\centering
\begin{tabular}{l|r} \hline
Central wavelength ($\lambda$) & 2.3 $\mu$m \\\hline
SPM parameter ($\gamma$) & 5.1 MW$^{-1}$ \\ \hline
Spectral filter bandwidth ($\Delta \lambda$) & 200 nm \\ \hline 
SAM parameter ($\kappa$) & 1 MW$^{-1}$ \\ \hline
SAM saturation parameter ($\zeta$) & 2 MW$^{-1}$ \\ \hline
Stiffness parameter ($\vartheta$) & 0.04 \\ \hline
Normalized energy ($E^\prime$) & $\kappa E_{cw}\sqrt{\zeta/\beta \gamma}$ \\ \hline
Normalized GDD parameter ($C$) & $\alpha \gamma/\beta \kappa$ \\ \hline
\end{tabular}
\caption{\label{tab:t1}Parameters of a Cr$^{2+}$:ZnS CPO.}
\end{table}

\section*{Results}
\label{sec:results}

\begin{figure}[ht]
\centering
\includegraphics[width=0.7\linewidth]{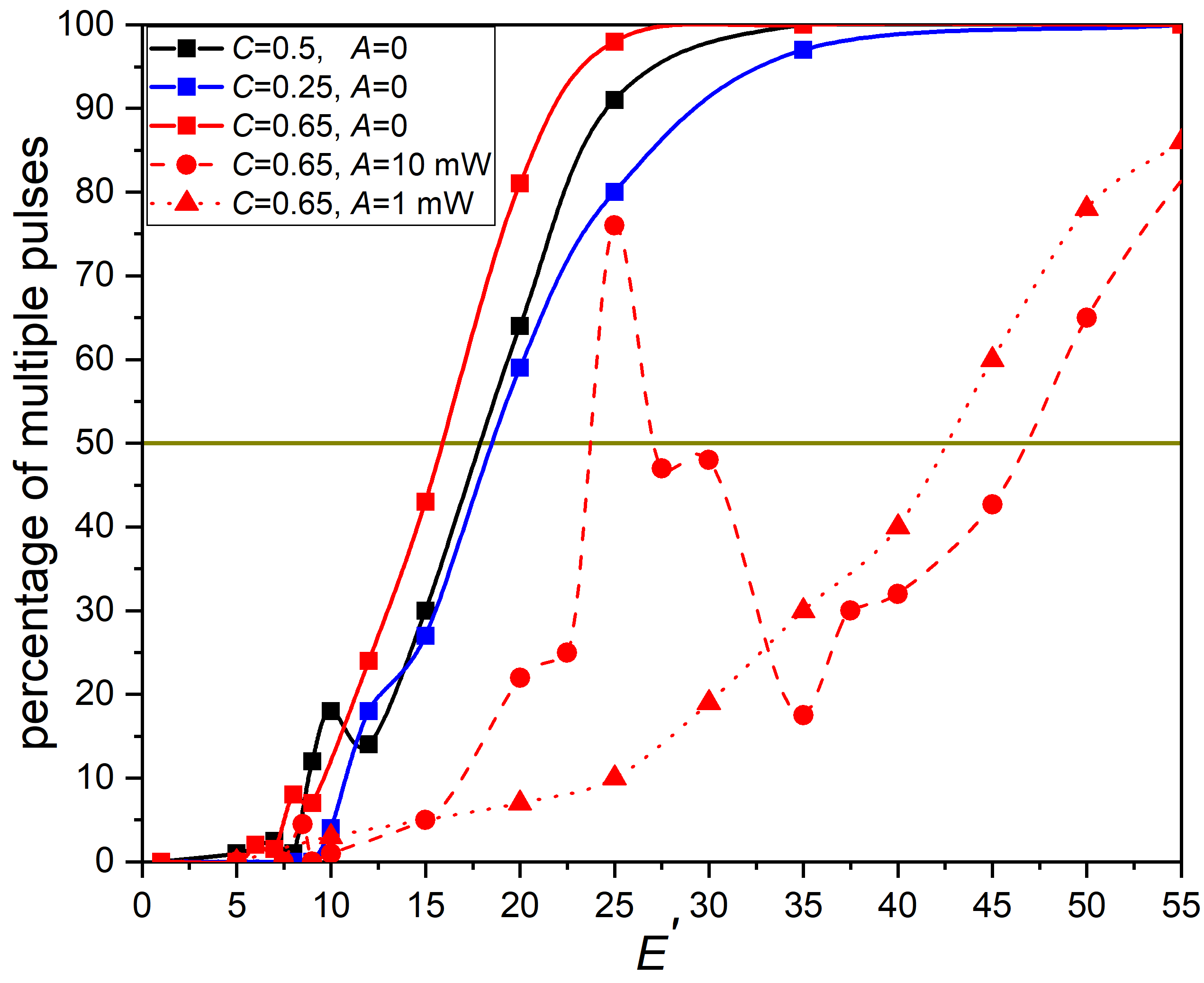}
\caption{The probabilities of multi-DS generation versus the dimensionless energy $E^\prime$ for different sets of ($C,A$). The physical parameters correspond to Table \ref{tab:t1}.}
\label{fig:fig1}
\end{figure}

We performed the statistics gathering over 100--200 independent samples for each set of parameters ($C$, $E^\prime$, $A$). The first two parameters were chosen because they define a parametric space of DS in the form of the ``master diagram'' and are used in the thermodynamic theory of DS stability \cite{kalashnikov2025energy}. We define the percentage counting of multipulse regimes that appear for a given parametric set as the primary characteristic of single DS stability (Fig. \ref{fig:fig1}).

In agreement with the thermodynamic theory of DS stability \cite{kalashnikov2025energy}, it can be characterized by the DS entropy and temperature under the condition of soliton energy quantization. When the soliton temperature reaches zero (or infinity, depending on the definition), the multi-soliton regime becomes more preferable. If $\mathcal{P}_1$ and $\mathcal{P}_2$ are the probabilities of single- and multi-soliton generation, then the effective temperature $T_{eff}$ could be defined from the following equation \cite{kramers1940brownian,RevModPhys.62.251,cugliandolo2011effective}:

\begin{align}
    \ln{(\frac{\mathcal{P}_2}{\mathcal{P}_1})}=-\frac{\Delta \Phi}{T_{eff}},\label{eq:temp}
\end{align}

\noindent where $\Delta \Phi$ is a quasipotential difference between two states 2 and 1. That is $T_{eff} \to \infty$ means $\mathcal{P}_2=\mathcal{P}_1$. This condition is indicated by the horizontal line in Fig. \ref{fig:fig1} and is defined by us as the threshold of single-DS destabilization \footnote{Of course, the contribution of $E^\prime$ to the DS energy landscape, i.e., $\Delta \Phi$, has to be taken into account. That makes the evolution of the probability distribution nontrivial, and we leave this issue for future consideration.}.

One can see that the stability threshold for $A=0$ lies between $E^\prime \approx 10 \div 20$, in agreement with the results of \cite{kalashnikov2025energy}, and grows slightly with the $C$-lowering. The transition to multi-pulsing with the energy growth is sufficiently sharp. Interestingly, the local picks of $\mathcal{P}_2$ appear within the stability zone, i.e., the zone is "grained." We interpret this phenomenon as a result of the interaction between the DS internal modes (see \cite{kalashnikov2025energy}) and quantum noise, i.e., a stochastic resonance \cite{gammaitoni1998stochastic}. The analysis of this phenomenon will be made elsewhere.

The situation changes drastically under the action of a single-mode low-power seed synchronized with the cavity round-trip (Fig. \ref{fig:fig1}, $A=$1 mW). The threshold of destabilization shifts to higher energies and becomes more gradual. Nevertheless, the further growth of seed energy ($A=$10 mW) changes this dependence dramatically. Instead of a gradual decrease in stability with energy, we observe a resonant enhancement of noise contribution within narrow energy intervals. That is an obvious manifestation of stochastic resonance when a weak regular signal amplifies the noise contribution and destabilizes DS. Moreover, such a signal suppresses mode-locking for low $E^\prime$ so that only the CW regime exists.

\begin{figure}[ht]
\centering
\includegraphics[width=0.9\linewidth]{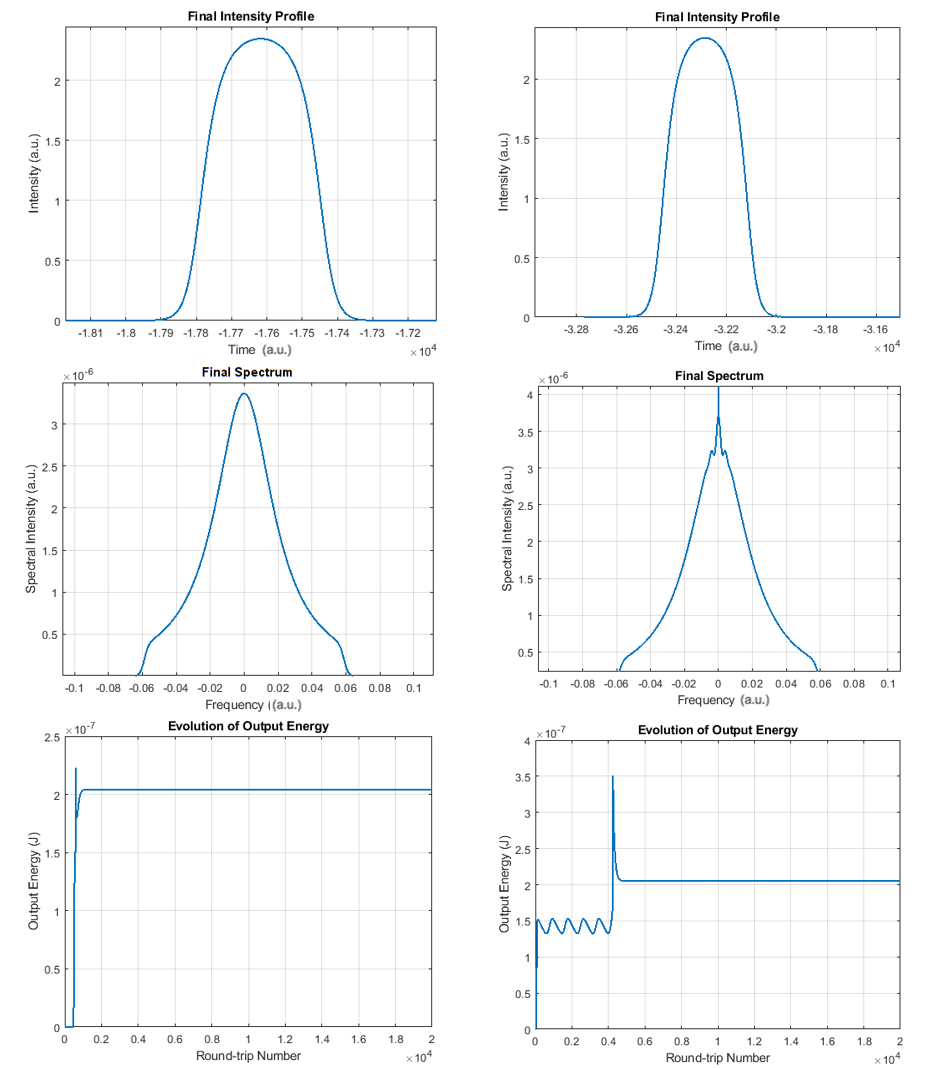}
\caption{The temporal (upper row), spectral (middle) profiles, and the single DS energy evolution (bottom). $A=$0 (left column) and 10 mW (right). $C=$0.65 and $E^\prime$=15.}
\label{fig:fig2}
\end{figure}

The last phenomenon can be interpreted based on Fig. \ref{fig:fig2}. The figure demonstrates the spectral and temporal profiles of DS as well as the evolution of its energy. The last demonstrates that the mode-locking forced by a seed becomes two-staged, i.e., DS oscillates before the mode-locking sustains (bottom-right picture). But the more radical difference is the appearance of an intense spectral spike at the central wavelength corresponding to the single-mode seed (middle-right picture, only part of the spike is shown). If a seed is sufficiently powerful, its contribution to CW can suppress mode-locking for small energies $E^\prime$. Simultaneously, its interaction with DS perturbs the spectrum at the central wavelength (see Figure), where the process of spectral condensation accompanies DSR (see \cite{kalashnikov2025energy}). As a result, the internal mode of DS can become disordered at some $E^\prime$.

\section*{Conclusion} \label{conc}

We have demonstrated that injecting a coherent, low-power single-mode seed into a noisy CPO provides a practical means to manage the DS stability. In the unseeded case, single-pulse operation is bounded by a sharp, threshold-like transition to multipulsing, consistent with a thermodynamic picture of strongly chirped pulses and their energy-scaling limits. Introducing a weak seed shifts this destabilization threshold to higher energies and makes the transition more gradual. Increasing the seed power, however, generates narrow “islands” of instability inside the nominal stability domain and can even suppress mode locking at low energies, in line with a resonance-assisted coupling between internal soliton modes and quantum noise of the basin. Our statistics-based analysis, utilizing probability maps in conjunction with energy and dispersion, identifies these islands and links them to mode–noise resonances triggered by the seed.  The conclusions are consistent with the thermodynamic theory of dissipative-soliton energy scalability, where the effective DS temperature and entropy jointly determine the stability of single- versus multi-pulse states.

Practically, our results suggest a control strategy for high-energy oscillators: employ a low-power seed to extend the usable single-pulse range, while avoiding seed–mode resonance windows indicated by the probability maps. Beyond lasers, the same mechanism should be observable in other driven dissipative platforms (e.g., microresonators and condensate analogs), offering a route to engineer stability by tailoring weak external drives.

\section*{Acknowledgments}

This work was supported by Norges Forskningsr{\aa}d (\#303347 (UNLOCK), \#326503 (MIR)), and ATLA Lasers AS. We acknowledge using the IDUN NTNU scientific cluster \cite{sjalander+:2019epic}.

\bibliography{sample}

\begin{thebibliography}{}

\bibitem[Battogtokh and Mikhailov, 1996]{battogtokh1996controlling}
Battogtokh, D. and Mikhailov, A. (1996).
\newblock Controlling turbulence in the complex ginzburg-landau equation.
\newblock {\em Physica D: Nonlinear Phenomena}, 90(1-2):84--95.

\bibitem[Chat{\'e} et~al., 1999]{chate1999forcing}
Chat{\'e}, H., Pikovsky, A., and Rudzick, O. (1999).
\newblock Forcing oscillatory media: phase kinks vs. synchronization.
\newblock {\em Physica D: Nonlinear Phenomena}, 131(1-4):17--30.

\bibitem[Chowdhury et~al., 2018]{chowdhury2018multipulse}
Chowdhury, S.~D., Pal, A., Chatterjee, S., Sen, R., and Pal, M. (2018).
\newblock Multipulse dynamics of dissipative soliton resonance in an all-normal
  dispersion mode-locked fiber laser.
\newblock {\em Journal of Lightwave Technology}, 36(24):5773--5779.

\bibitem[Cross and Hohenberg, 1993]{cross1993pattern}
Cross, M.~C. and Hohenberg, P.~C. (1993).
\newblock Pattern formation outside of equilibrium.
\newblock {\em Reviews of modern physics}, 65(3):851.

\bibitem[Cugliandolo, 2011]{cugliandolo2011effective}
Cugliandolo, L.~F. (2011).
\newblock The effective temperature.
\newblock {\em Journal of Physics A: Mathematical and Theoretical},
  44(48):483001.

\bibitem[Gammaitoni et~al., 1998a]{RevModPhys.70.223}
Gammaitoni, L., H\"anggi, P., Jung, P., and Marchesoni, F. (1998a).
\newblock Stochastic resonance.
\newblock {\em Rev. Mod. Phys.}, 70:223--287.

\bibitem[Gammaitoni et~al., 1998b]{gammaitoni1998stochastic}
Gammaitoni, L., H{\"a}nggi, P., Jung, P., and Marchesoni, F. (1998b).
\newblock Stochastic resonance.
\newblock {\em Reviews of modern physics}, 70(1):223.

\bibitem[Goyal et~al., 2012]{goyal2012lorentzian}
Goyal, A., Raju, T.~S., Kumar, C.~N., et~al. (2012).
\newblock Lorentzian-type soliton solutions of ac-driven complex
  ginzburg--landau equation.
\newblock {\em Applied Mathematics and Computation}, 218(24):11931--11937.

\bibitem[Grelu and Akhmediev, 2012]{grelu2012dissipative}
Grelu, P. and Akhmediev, N. (2012).
\newblock Dissipative solitons for mode-locked lasers.
\newblock {\em Nature photonics}, 6(2):84--92.

\bibitem[Grelu et~al., 2010]{grelu2010dissipative}
Grelu, P., Chang, W., Ankiewicz, A., Soto-Crespo, J.~M., and Akhmediev, N.
  (2010).
\newblock Dissipative soliton resonance as a guideline for high-energy pulse
  laser oscillators.
\newblock {\em JOSA B}, 27(11):2336--2341.

\bibitem[H\"anggi et~al., 1990]{RevModPhys.62.251}
H\"anggi, P., Talkner, P., and Borkovec, M. (1990).
\newblock Reaction-rate theory: fifty years after kramers.
\newblock {\em Rev. Mod. Phys.}, 62:251--341.

\bibitem[Haus and Mecozzi, 1993]{206583}
Haus, H. and Mecozzi, A. (1993).
\newblock Noise of mode-locked lasers.
\newblock {\em IEEE Journal of Quantum Electronics}, 29(3):983--996.

\bibitem[Kalashnikov, 2017]{kalashnikov2017chaotic}
Kalashnikov, V.~L. (2017).
\newblock Chaotic, stochastic resonance, and anti-resonance phenomena in
  optics.
\newblock {\em Resonance}, 4:70--87.

\bibitem[Kalashnikov et~al., 2025]{kalashnikov2025energy}
Kalashnikov, V.~L., Rudenkov, A., Sorokin, E., and Sorokina, I.~T. (2025).
\newblock Energy scalability limits of dissipative solitons.
\newblock {\em Physical Review A}, 111(4):043529.

\bibitem[Kippenberg et~al., 2018]{doi:10.1126/science.aan8083}
Kippenberg, T.~J., Gaeta, A.~L., Lipson, M., and Gorodetsky, M.~L. (2018).
\newblock Dissipative kerr solitons in optical microresonators.
\newblock {\em Science}, 361(6402):eaan8083.

\bibitem[Komarov et~al., 2013]{komarov2013competition}
Komarov, A., Amrani, F., Dmitriev, A., Komarov, K., and Sanchez, F. (2013).
\newblock Competition and coexistence of ultrashort pulses in passive
  mode-locked lasers under dissipative-soliton-resonance conditions.
\newblock {\em Physical Review A—Atomic, Molecular, and Optical Physics},
  87(2):023838.

\bibitem[Korobko et~al., 2023]{korobko2023birth}
Korobko, D., Ribenek, V., Itrin, P., and Fotiadi, A. (2023).
\newblock Birth and annihilation of solitons in harmonically mode-locked fiber
  laser cavity through continuous wave injection.
\newblock {\em Optical Fiber Technology}, 75:103216.

\bibitem[Kramers, 1940]{kramers1940brownian}
Kramers, H.~A. (1940).
\newblock Brownian motion in a field of force and the diffusion model of
  chemical reactions.
\newblock {\em physica}, 7(4):284--304.

\bibitem[Malomed and Nepomnyashchy, 1990]{malomed1990kinks}
Malomed, B.~A. and Nepomnyashchy, A.~A. (1990).
\newblock Kinks and solitons in the generalized ginzburg-landau equation.
\newblock {\em Physical Review A}, 42(10):6009.

\bibitem[Rudenkov et~al., 2023]{Rudenkov}
Rudenkov, A., Kalashnikov, V.~L., Sorokin, E., Demesh, M., and Sorokina, I.~T.
  (2023).
\newblock High peak power and energy scaling in the mid-ir chirped-pulse
  oscillator-amplifier laser systems.
\newblock {\em Opt. Express}, 31(11):17820--17835.

\bibitem[Sj\"alander et~al., 2019]{sjalander+:2019epic}
Sj\"alander, M., Jahre, M., Tufte, G., and Reissmann, N. (2019).
\newblock {EPIC}: An energy-efficient, high-performance {GPGPU} computing
  research infrastructure.

\end{thebibliography}

\end{document}